# Low-frequency spins and the ground state in high-$T_c$ cuprates

C. Panagopoulos[a*], J.L. Tallon[b], B.D. Rainford[c], J.R. Cooper[a], C.A Scott[c] and T. Xiang[d]

[a] Cavendish Laboratory and IRC in Superconductivity, University of Cambridge, Cambridge CB3 0HE, United Kingdom
[b] New Zealand Institute for Industrial Research, P.O. Box 31310, Lower Hutt, New Zealand
[c] Department of Physics and Astronomy, University of Southampton, Southampton S017 1BJ, United Kingdom
[d] Institute of Theoretical Physics, Academia Sinica, P.O. Box 2735, Beijing 100080, Peoples Republic of China

## Abstract

We study the spin fluctuations of pure and Zn-substituted high temperature superconductors (HTS) using the muon spin relaxation ($\mu$SR) technique. Superconductivity is found to coexist with low frequency spin fluctuations over a large region of the superconducting phase diagram. The characteristic temperature of spin fluctuations detected by $\mu$SR decreases with increasing doping and vanishes above a critical doping $p_c \approx 0.19$ where the normal state pseudogap vanishes. Furthermore, it is at $p_c$ that our penetration depth studies show an abrupt change in the doping dependence of the superfluid density, $\rho_s$. For $p > p_c$ the absolute value of $\rho_s$ is large and nearly independent of carrier concentration whereas for $p < p_c$ it drops rapidly, signalling a crossover to weak superconductivity. These results are discussed in terms of a quantum transition involving fluctuating short-range magnetic order that separates the superconducting phase diagram of HTS into two distinct ground states.

## 1. Introduction

The discovery of superconductivity in layered copper-oxide compounds came as a great surprise, not only because of the record high $T_c$'s, but also because these materials are relatively poor metals even at high doping levels. Superconductivity in high-$T_c$ superconductors (HTS) is obtained by electronically doping parent compounds that are antiferromagnetic (AF) Mott insulators - materials in which both the AF and the insulating behaviour are the result of strong electron-electron interactions. Because local magnetic correlations survive in the doped metallic compounds, we can view these materials as doped antiferromagnets.

In HTS at low temperature it is generally accepted that long-range AF order is suppressed with the addition of the first few added holes and the ground-state transforms to a short-range ordered spin-glass-state. At an apparently universal doping state of $p$~0.05 holes/planar-Cu-atom superconductivity emerges, in coexistence with low energy spin fluctuations. With further doping correlations gradually weaken and these materials progress to a more Fermi-liquid-like state with the eventual demise of superconductivity in the neighbourhood of $p$~0.27. This seemingly gradual evolution, poses important questions as to the nature of the ground state especially the possible presence of magnetic domains, intrinsic or extrinsic electronic inhomogeneity and the nature and role of the pseudogap.

Although, for the most part, these materials evolve gradually there is mounting experimental evidence for quantum critical behaviour near one or more points of the $T$-$p$ phase diagram [1,2]. The two points of most interest are $p$=0.125 and $p$=0.19. $^{63}$Cu nuclear quadrapole resonance measurements find that the charge stripe order parameter seems to vanish for $p$>1/8 although the experimental error does not allow definite conclusions to be drawn [3]. The second critical doping point has received greater attention mainly because it is at $p$=0.19 that one may observe significant changes in the behaviour of the superconducting and normal state properties. Fundamental physical quantities such as the superconducting condensation energy [2], the superfluid density [4,5], quasiparticle weight [6] and resistivity power law [7] all show distinct changes at the critical doping $p_c$=0.19. It is also at $p_c$ that the ("upper") pseudogap characteristic temperature, $T^*$, determined from heat capacity, Knight shift and resistivity extrapolates to zero [2]. A further "lower" pseudogap temperature $T^{**}$ has been suggested from angular resolved photoemission spectroscopy (ARPES) and scanning tunnelling microscopy. At $p=p_c$=0.19, $T^{**}$ seems to equal $T_c$ [8]. Whether $T^*$ and $T^{**}$ are related and whether either or both have anything to do with superconductivity is yet to be clarified.

Significant effort has been expended in attempting to identify a quantum transition in HTS. Thermodynamic studies show no clear evidence of a phase transition in the normal state. However, below $T^*$ the entropy is significantly suppressed reflecting developing correlations in the normal state, and there is mounting evidence in support of a relation between these thermodynamic changes and the evolution of low frequency spin fluctuations [1,2]. This is certainly not surprising given, as discussed above, the way these materials evolve with the addition of carriers. Yet there is still no definite experimental investigation addressing these apparently fundamental links.

## 2. Experimental

A technique extensively used to study the low energy ($10^{-9} - 10^{-6}$ s) spin dynamics in magnetic materials and alloys is muon spin relaxation ($\mu$SR) [9]. It has also been successfully employed in the study of HTS and indeed provided some of the earliest evidence for the presence of low energy spin excitations in various HTS [10-15].

It is therefore potentially very instructive to perform a systematic $\mu$SR investigation of the low-energy spin dynamics across the $T$-$p$ phase diagram in several HTS and preferably in small increments in doping state. Moreover, it is necessary to correlate the results with the behaviour of other fundamental physical properties reflecting the superconducting ground state. To this aim we have synthesised samples of $La_{2-x}Sr_xCu_{1-y}Zn_yO_4$ ($x$=0.03-0.24 and $y$=0, 0.01 and 0.02) using solid-state reaction followed by quenching and subsequent oxygenation. We note that in $La_{2-x}Sr_xCuO_4$ (La-214) $x=p$. The samples were characterised by powder x-ray diffraction, micro-Raman spectroscopy as well as extensive transport and thermodynamic measurements and found to be phase pure. Their $T_c$ values (Fig. 1) agree with published data for powders and single crystals [16,17]. The lattice parameters were also in good agreement with published data, where available [16]. Care was taken to ensure high purity and homogeneity. For example, we dry all precursor powders; we react, grind, mill, re-press and re-sinter at least 4 times to ensure extrinsic homogeneity. In the Zn-doped samples we encapsulate to ensure minimal Zn loss. To avoid possible phase separation we quench from the last high temperature reaction then begin our oxygen anneal at 750°C and apply a slow cool over many days to 350°C to ensure full oxygenation of vacant oxygen sites associated with Sr substitution. The heat capacity anomalies in this material are sharp, the resistive and $ac$-susceptibility transitions are sharp, and micro-analytical spectroscopic studies show no signature of impurities or inhomogeneity. We also note that heat capacity data from our group shows a similarly-sized-jump in the electronic specific heat coefficient for $p$=0.22 and $p$=0.24 [18]. This indicates superconductivity is a bulk property even for $p$=0.24. In particular for Sr~0.20, where we observe abrupt changes in the spin and charge excitation spectrum inhomogeneity would tend to round out these abrupt changes. This would imply that any changes we report below at $p$~0.19 may in fact be even more abrupt than we show.

Zero-field (ZF) and transverse-field (TF) $\mu$SR studies were performed at the pulsed muon source, ISIS Facility, Rutherford Appleton Laboratory. Spectra were collected down to as low as 40mK thus allowing the temperature dependence of slow spin fluctuations to be studied to high doping. In a $\mu$SR experiment, 100% spin-polarised positive muons implanted into a specimen precess in their local magnetic environment. Random spin fluctuations will depolarise the muons provided they do not fluctuate much faster than the muon precession. The muon decays with a life-time of 2.2$\mu$s, emitting a positron preferentially in the direction of the muon spin at the time of decay. By accumulating time histograms of such positrons one may deduce the muon depolarisation rate as a function of time after implantation. The muon is expected to reside at the most electronegative site of the lattice, in this case 1.0Å away from the apical $O^{2-}$ site above and below the $CuO_2$ plane so the results reported here are dominated by the magnetic correlations in the $CuO_2$ planes.

To investigate the evolution of the ground state we measured perhaps the most fundamental quantity in superconductivity, namely the superfluid density. The superfluid density, $\rho_s \sim \lambda_{ab}^{-2}$, was determined from measurements of the in-plane magnetic penetration depth $\lambda_{ab}$ using the low-field $ac$-susceptibility technique at 1G (parallel to the $c$-axis) and 333Hz for grain-aligned powders. To magnetically align the samples, the powder was first lightly ground by hand to remove aggregates and passed through a sieve. In La-214 it is common to obtain small grain agglomerates of the order of a few microns. Randomly oriented grain agglomerates can be a cause of poor alignment, and to eliminate these, powders obtained after sieving were ball milled in ethanol and dried after adding a defloculant. Scanning electron microscopy confirmed the absence of grain boundaries and showed that the grains were approximately spherical with grain diameter ~4 $\mu$m. The powders were mixed with a 5-min curing epoxy and aligned in a static field of 12 T at room temperature. Debye-Scherrer x-ray scans showed that approximately 90% of the grains had their $CuO_2$ planes aligned to within approximately 1.8 degrees. The $ac$-susceptibility measurements were performed down to 1.2K with a home-made susceptometer using miniature coils with the $ac$ field $H_{ac}$ applied along the $c$-axis. The Earth's field was screened out using a mu-metal shield. Measurements were performed for $H_{ac}$=1G rms at $f$=333Hz. The separation of the grains and the absence of weak links were confirmed by checking the linearity of the signal for $H_{ac}$ from 0.3 to 3G rms and $f$ from 33 to 333Hz. We also used a commercial susceptometer to confirm some of our findings and check the linearity up to 10G rms. Taking the grains to be approximately spherical, as indicated by scanning electron microscopy, the data were analysed using London's model [19,20], and the absolute values and variation of $\lambda_{ab}$ with temperature were obtained. In total, 16 samples were investigated for each doping content. Furthermore, the values of $\lambda_{ab}^{-2}(0)$ for some of the samples were also confirmed by standard transverse field $\mu$SR experiments performed on unaligned powders at 400G [21].

3. **Results and Discussion**

Figure 2 shows the typical time dependence at several temperatures of the ZF muon asymmetry for pure $La_{2-x}Sr_xCuO_4$ with $x$=$p$=0.08. In all samples the high-temperature form of the depolarisation is Gaussian and temperature independent, consistent with dipolar interactions between the muons and their near-neighbour nuclear moments. This was verified by applying a 50G longitudinal field, which completely suppressed the depolarisation. Here the electronic spins in the $CuO_2$ planes fluctuate so fast that they do not affect the muon polarisation.

On cooling the fluctuations slow until they enter the $\mu$SR time window and begin to contribute to the relaxation of the polarisation. This is clearly evident in the figure with the crossover from negative to positive curvature in the time-decay of the asymmetry. At low enough temperatures the fluctuations freeze out and, typical of other spin glass systems, there is a fast relaxation due to a static distribution of random local fields, followed by a long-time tail with a slower relaxation resulting from remnant dynamical processes within the spin glass.

To study the doping dependence of this slowing down we determine two characteristic temperatures. (i) The temperature, $T_f$, where the spin correlations first enter the $\mu$SR time window i.e., where the muon asymmetry first deviates from Gaussian behaviour and (ii) the temperature, $T_g$, at which these correlations freeze into a glassy state thus causing an initial rapid decay in the asymmetry.

$\mu$SR is sensitive to spin fluctuations within a time window of $10^{-9}$s to $10^{-6}$s and we may therefore associate $T_f$ and $T_g$, respectively, with these lower and upper thresholds. In general the relaxation data may be fitted to the form $G_z(t)=A_1\exp(-\gamma_1 t)+A_2\exp(-(\gamma_2 t)^\beta)+A_3$ where the first term is the fast relaxation in the glassy state (i.e., at higher temperatures $A_1=0$), the second "stretched-exponential" term is the slower dynamical term and $A_3$ accounts for a small time-independent background arising from muons stopping in the silver backing plate. As in other spin glass systems, in the high-temperature Gaussian limit $\beta=2$ [9-15,22]. Consequently, any departure below $\beta=2.0\pm0.06$ (Fig. 3) is taken as the onset temperature, $T_f$, at which spin fluctuations slow down sufficiently to enter the time scale of the muon probe ($10^{-9}$s).

As a further check on the assignment of $T_f$ we fitted the high-temperature data to the full Kubo-Toyabe function $G_z(t)=A_1\exp(-\alpha t^2)\exp(-\gamma t)+A_2$. The relaxation rate $\gamma$ is found to rise from zero at the same temperature at which the exponent $\beta$ departs from 2, confirming the entrance of the spin correlations into the experimental time window [23].

At low temperatures the exponent $\beta$ falls rapidly towards the value 0.5 (Fig. 3) as expected for a spin glass [22]. We identify $T_g$ as the temperature at which $\beta=0.5\pm0.06$. This "root exponential" form for the relaxation function is a common feature of spin glasses, and in the present samples the temperature $T_g$ coincided with a maximum in the longitudinal relaxation rate $\gamma$ [23] and the appearance of the fast relaxation. Other methods of analysis may still be possible and a different choice might affect the magnitude of the $T_g$ values but not the trends. Our values for $T_g$ agree with published data obtained by different techniques, where available [10-15,24-26].

We first discuss the data for pure La-214 (i.e., $y=0$ in Fig. 4). Also included (Sr<0.03) is data from other independent studies [13]. With the first added holes we see below $T_g$ the development of short-range magnetic order, presumably associated with interaction between moments in the doped oxygen orbitals. With increasing doping $T_g$ is found to increase until the Neel temperature vanishes [13,27]. For $p>0.02$ the freezing temperature, $T_g$, of this short-range magnetic phase then decreases with further doping. This behaviour is common in HTS (pure $YBa_2Cu_3O_{7-\delta}$ (Y-123), $Y_{1-y}Ca_yBa_2Cu_3O_{6.02}$ (Ca:Y-123) and La-214) and has been seen both by neutron scattering and $\mu$SR [10-15,25-28]. Moreover, we find that although the freezing occurs at very low temperatures, $T_g$, low-frequency spin fluctuations enter the experimental time window at significantly higher temperatures, $T_f$. Values of $T_g$ and $T_f$ summarised in Fig. 4 indicate that the spin-glass phase persists beyond $p=0.125$. In fact the onset of the spin glass phase for $p=0.125$ occurs at a higher temperature than that for $p=0.10$. This may be due to the formation of strongly-correlated antiferromagnetic stripe domains [29-32] which have been inferred in this range of doping from inelastic neutron scattering (INS) studies [29]. We also note that although the doping dependence of the glass transition temperature $T_g$ does not seem to change much with the onset of superconductivity (at $p \sim 006$), $T_f$ does. This observation is not surprising in view of the strong modification of the spin spectrum accompanying the onset of superconductivity and seen e.g. in INS by the growth of the magnetic resonance [17]. For $p=0.15$ and 0.17, $T_g$ becomes very small (<45mK) and $T_f$ is approximately 8K and 2K, respectively. For $p\geq0.20$, there are no changes in the depolarisation function to the lowest temperature measured (40mK) suggesting that here the AF spin fluctuations have very short lifetimes, certainly outside of the $\mu$SR time window. Similar observations have been made in $La_{1.6-x}Nd_{0.4}Sr_xCuO_4$ single crystals where no magnetic order was observed for $p\geq0.20$ [14].

It may be argued that dynamic magnetic effects may not be conclusively distinguished from static magnetism using the zero-field line shape alone. We can say that for $p \leq 0.08$ oscillations in the asymmetry were observed and by decoupling experiments in a longitudinal field, we confirmed the static nature of the magnetic ground state there. For $p > 0.08$ oscillations were not observed and the data was better represented by an exponential relaxation indicating either a very strongly disordered static field distribution or rapid fluctuations. Moreover we did not rely exclusively on the slow depolarisation but noted (see above) the presence of an additional fast relaxation occurring exclusively below $T_g$ and consistent with disordered static dipoles. This is not seen beyond $p=0.19$. We therefore believe the case for a crossover from slow dynamic fluctuations to static short-range order to be strong.

It is tempting to infer from the absence of dynamic effects ($T_f=0$) that AF spin fluctuations are no longer present for $p>0.19$. Given the frequency limitation of the $\mu$SR technique one can only say that these spins fluctuate with a lifetime $< 10^{-9}$s. Also our failure to observe $T_g$ for $p>0.15$ in the La-214 series (Fig. 4) again may simply be due to the limits set by the measurement technique. In fact, as shown in the inset of Fig. 4, $T_g$ and $T_f$ scale rather well suggesting a general systematic slowing of the AF spin fluctuations about $p_c$. This might be taken as an indication that an observable $T_f$ signals the presence of an observable $T_g$ yet the latter is subject to limitations of the measurement tool. This is a crucial question, which will be partially addressed below in our Zn substitution studies but probably deserves further close investigation. Irrespective of the relation between $T_f$ and $T_g$ it is important to check whether $p=0.19$ is robust with respect to the energy window or is clearly a number associated with the frequency limit of the technique. To this aim we need to either slow down the spins or expand the frequency window of the technique. In our experiment it is perhaps easier to do the former and this is the route we have followed.

Earlier spectroscopic studies, including INS [17,33] and nuclear magnetic resonance [34] experiments, show that Zn substitution has a two-fold impact in slowing spin fluctuations and suppressing long-range magnetic order. This would imply that Zn doping should promote spin-glass behaviour. Indeed, as depicted in Fig. 5, Zn enhances the muon depolarisation rate at low temperatures and causes an increase in both $T_g$ and $T_f$. The striking result which Fig. 5 summarises is the apparent convergence of both $T_g(p)$ and $T_f(p)$ to zero, for all Zn concentrations, at the critical doping $p_c \approx 0.19$. While this effect is not so obvious for the pure samples it is very clear in the two Zn-substituted series. The fact that $T_f(p) \rightarrow 0$ as $p \rightarrow p_c$ for all Zn concentrations suggests that spin correlations within the upper $\mu$SR time threshold of $10^{-9}$s die out beyond $p_c$ leaving only very short-lived fluctuations (or none at all) beyond $p_c$. The fact that $T_f$ and $T_g$ both vanish as $p \rightarrow p_c$ implies that the rate of slowing down diverges at $p_c$, in the sense that *the characteristic time changes from $10^{-9}$ and $10^{-6}$s in progressively smaller temperature intervals as $p_c$ is approached*. In the absence of evidence for long-range order in the normal state, the present observations indicate the existence of a quantum glass transition at $p_c$ as signalled by the divergence in the rate of slowing down at $p_c$.

Therefore, both $T_g$ and $T_f$ are found to decrease with increasing doping and go to zero at $p \approx 0.19$. Furthermore, we can state categorically that the present results are not peculiar just to La-214. La-214 is sometimes said to be atypical of HTS because of its propensity for "stripe" behaviour and its apparently high degree of inhomogeneity. Yet we observe similar behaviour in the $Bi_{2.1}Sr_{1.9}Ca_{1-x}Y_xCu_2O_{8+\delta}$ (Bi-2212) system. (In our samples we used $x=0$, 0.3, 0.5 and appropriate values of $\delta$ to achieve the desired carrier concentration. The optimally doped sample had $x=0$.) Underdoped samples were prepared by deoxygenation and quenching (note that deoxygenation actually removes disorder in Bi-2212). The samples were fully characterised using also thermoelectric power to determine the doping state (Fig. 6). Figure 7, shows our data for this material overlaid on the La-214 data. Bi-2212 shows precisely the same trend with $T_g$ and $T_f \rightarrow 0$ as the hole concentration $p \rightarrow 0.19$. We note further that the doping dependence of $T_g$ seen here has also been found in pure Y-123 [28] and in Ca:Y-123 up to 0.09 holes per planar copper atom [13] suggesting that the behaviour shown in Fig. 7 is generic and common to all high-$T_c$ materials. This is not to discount significant effects due to stripes in La-214. Indeed such effects are clearly seen in the figure where the anomaly seen in La-214 at $p=0.125$ is clearly absent or diminished in the Bi-2212 data. Also, the glassy behaviour would seem to not be associated with the degree of material disorder. Values of $T_g$ for the Y-123 and Bi-2212 families, believed to pose less disorder than La-214, are in fact twice the values of the latter. At the same time Y-123 is generally regarded as a much cleaner system than Bi-2212. Similarly, $T_g$ for pure Y-123 is about the same as $T_g$ for Ca:Y-123 in spite of the substantial disorder in the latter system [13,28].

It is perhaps of interest that the temperature scales $T_g$, $T_f$ and their doping dependence are similar to those of the pseudogap temperature $T^*$ which also vanishes at $p\sim0.19$, coincident with the glass transition (Fig. 7). Taken together, these results provide evidence for coexistence of low energy spin fluctuations with superconductivity, as well as a link between the identified magnetic scales and $T^*$. In principle, it seems that by increasing the Zn concentration it may be possible to continue pushing the $T_g$ and $T_f$ values to higher temperatures and, at the same time, further suppress superconductivity. This should motivate further searches for a possible direct correlation between the pseudogap and the magnetic ground state. Such studies could specifically address the following two questions. First, as to whether $T^*$ is the asymptote of $T_f$. Second if the absence of dynamical relaxation for $p>p_c$ suggested by our present results is robust to the energy scale of spin fluctuations, hence the presence of an unambiguous quantum transition. These interesting and important questions are the subject of our present investigations.

We now turn to the penetration depth measurements. We have obtained systematic results on the effects of carrier concentration on the superfluid density of La-214 (Fig. 8). In the overdoped region we find a reasonably constant value of $\rho_s(0)$ and the temperature dependence is in good agreement with the weak-coupling $d$-wave temperature dependence. (In Fig. 8 $\rho_s(T)$ for $p=0.24$ still shows significant deviations from the $d$-wave curve that possibly reflect changes in the electronic structure. In fact the data for $p=0.24$ are in excellent agreement with a weak-coupling $d$-wave calculation for a rectangular Fermi surface [35]. This would not be surprising given the changes in Fermi surface with the rapid crossover from hole-like to electron-like states in that region, as observed by ARPES experiments [36].) In the optimal and underdoped regions $\rho_s(0)$ is rapidly suppressed and there is a marked departure of the temperature dependence of $\rho_s(T)$ from the canonical weak coupling curve. Regarding the accuracy of the data we note that each data point represents a total of 16 samples. Given that all our samples were prepared under the same conditions, the size and shape of the grains were essentially the same for all Sr concentrations and the $\lambda_{ab}(0)$ values measured by both the ac-susceptibility and $\mu$SR techniques are in excellent agreement, we believe the actual error is significantly lower than the estimates shown in Fig. 8.

It is important to note that as for $\lambda_{ab}$ we have also observed a similar behaviour in the $c$-axis component. In Fig. 9 we show our $\lambda_c^{-2}(0)$ data for La-214 and HgBa$_2$CuO$_{4+\delta}$ (Hg-1201) [37]. Both HTS show a clear drop in $\lambda_c^{-2}(0)$ near $p_c$, with Hg-1201 displaying the sharpest change of all HTS to date, which could be related to the high degree of homogeneity and order in this monolayer cuprate. We also note in La-214, where we have measured more dopings than in Hg-1201, that the increase of $\lambda_c^{-2}(0)$ with doping in the region $p=0.05$ to $\sim0.20$ is not linear but instead follows a power law of approximately 2.7. This it self is an interesting result which deserves further investigation.

The competition between quasi-static magnetic correlations and superconductivity thus results in a marked suppression of the superfluid density. This suppression in the underdoped region, referred to as weak superconductivity, has been directly linked to the strong reduction in entropy and condensation energy associated with the pseudogap [4]. The direct relationship between the doping dependence of the superfluid density and the electronic entropy may be understood within a Fermi surface model. In this model the entropy may be expressed as the average of the product of the Fermi velocity, $v_F$, and the density of states over an energy interval $E_F \pm \Delta$, where $\Delta$ is the superconducting gap [4,18]. In fact $v_F$ does not appear to alter much with doping [38] thus ensuring a direct link between the entropy at $T_c$ and the superfluid density.

We note the abrupt drop in $\rho_s(0)$ near 0.19. A similar behaviour has now been observed in Bi-2212 [39] and was previously reported for Ca:Y-123 and Tl$_{0.5-y}$Pb$_{0.5+y}$Sr$_2$Ca$_{1-x}$Y$_x$Cu$_2$O$_7$ [5]. The observed abrupt collapse of the superconducting "strength" casts doubt on the interpretation of the anomalous properties of HTS in terms of doping inhomogeneity, as suggested by recent tunnelling microscopy studies [40]. The relatively doping independent overdoped region in La-214 and Bi-2212 [39] seems to differ from data for some other HTS. Unlike La-214 and Bi-2212 each of Tl$_2$Ba$_2$CuO$_{6+\delta}$ [41,42], Ca:Y-123 and Tl$_{0.5-y}$Pb$_{0.5+y}$Sr$_2$Ca$_{1-x}$Y$_x$Cu$_2$O$_7$ [5] show a drop in the superfluid density in the overdoped regime. The origins of this difference are not clear and remain an open and important question, which calls for further investigation. Nevertheless, the maximum of $\rho_s$ at $p_c$ is undoubtedly universal.

The detailed ac-susceptibility data for the penetration depth serve the important role in identifying directly the interplay between magnetism and superconductivity and in revealing a close relationship between the onset of slow spin fluctuations (and spin glass behaviour at low temperature) and the observed sharp reduction in superfluid density at $p_c$ =0.19. This shows, in particular, that the onset of short-range magnetic correlations coincides with an abrupt change in the superconducting ground state. This is a robust result which has now been established for both La-214 and Bi-2212 [23,39].

Tallon and Loram have compiled an extensive set of experimental data that demonstrates "special behaviour at $p$=0.19" [2]. From thermodynamic data they deduced the doping dependence of the density of states and in particular the abrupt disappearance of a quasiparticle gap (the pseudogap) at $p$=0.19. They supported this picture by considering in addition a number of other spectroscopies. Irrespective of how this data might be interpreted it suggested that important changes occur in the quasiparticle spectrum at $p$=0.19. Here we provide experimental data showing simultaneous abrupt changes in the magnetic spectrum at $p$=0.19 namely, a similarly sudden disappearance of low-frequency spin fluctuations. The picture that emerges is that quasiparticle lifetimes at ($\pi$, 0) are strongly damped for $p$<0.19 and are suddenly recovered beyond while spin fluctuations are strongly damped for $p$>0.19 [43]. We suggest that the observation of simultaneous abrupt changes in the magnetic and quasiparticle spectrum is a pivotal new result that places strong constraints on theoretical models.

Aspects of our work complement a number of other studies which emphasise links between superconductivity and magnetism [44-50]. In fact, such links have been assumed since the earliest discoveries in HTS. This work provides a context for our study. However, the detailed nature of the link between superconductivity and magnetism has never been clarified. Our observation of a quantum glass transition suggests a possible central role of quantum-critical magnetic fluctuations and associated dynamical crossovers.

It has become increasingly evident that the long-standing mystery of HTS will not be resolved until some new "missing element" is identified. While it has been recognised for several years now that fluctuations near a quantum transition could possibly explain both the origin of superconductivity in HTS and their unusual normal-state properties, no clear evidence for such has been presented. To illustrate we quote from the recent article by Si *et al*. [51]: "The extensive present interest in metals close to a second-order quantum phase transition has stemmed largely from studies of high-temperature superconductors. In these systems, however, it has been hard to locate the putative quantum critical points". Specifically (i) no long-range ordered state has been found, (ii) no thermodynamic transition has been found and (iii) no evidence for critical slowing down has been found. Here we have presented evidence in support of a quantum transition involving fluctuating short-range magnetic order which crosses over discontinuously at $T$=0 to quasi-static behaviour. This means that (i) and (ii) no longer need be satisfied and we have now demonstrated slowing down of spin fluctuations near critical doping. It seems that such a quantum phase transition divides the phase diagram of HTS in two distinct regions: one of "weak" superconductivity where the latter coexists and competes with 2D spin fluctuations, and a region of "strong" BCS-like superconductivity where the short-range spin correlations seem to be absent.

## Acknowledgements

C.P. acknowledges the support of The Royal Society (London). J.L.T. acknowledges financial assistance from the New Zealand Marsden Fund and T.X. from the National Natural Science Foundation of China.

## References

1. S. Sachdev *Physics World* **12** (1999), p. 33; S. Sachdev *Science* **288** (2000), p. 475.
2. J.L. Tallon and J.W. Loram *Physica C* **349** (2001), p.53.
3. A.W. Hunt *et al*. *Phys. Rev. Lett*. **82** (1999), p. 4300.
4. C. Panagopoulos *et al*. *Phys. Rev. B* **60** (1999), p. 4617.
5. C. Bernhard *et al*. *Phys. Rev. Lett*. **86** (2001), p. 1614.
6. D.L. Feng *et al*. *Science* **289** (2000), p. 277.
7. S. Naqib, J.R. Cooper , J.L. Tallon and C. Panagopoulos *preprint*.
8. T. Nakano, N. Momono, M. Oda and M. Ido *J Phys. Soc. Jpn.* **67** (1998), p. 2622.
9. Y.J. Uemura, T. Yamazaki, D.R. Harshman, M. Senba and E.J. Ansaldo *Phys. Rev. B* **31** (1985), p. 546.
10. D. R. Harshman *et al*. *Phys. Rev. B* **38** (1988), p. 852.
11. J. I. Budnick *et al*. *Europhys. Lett*. **5** (1988), p. 65.
12. R.F. Kiefl *et al*. *Phys. Rev. Lett*. **63** (1989), p. 2136.
13. Ch. Niedermayer *et al*. *Phys. Rev. Lett.* **80** (1998), p. 3843.
14. B. Nachumi *et al*. *Phys. Rev. B* **58** (1998), p. 876.

15. A. Kanigel, A. Keren, Y. Eckstein, A. Knizhnik, J.S. Lord and A. Amato *Phys. Rev. Lett.* **88**, (2002), p. 137003.
16. P.G. Radaelli *et al. Phys. Rev. B* **49** (1994), p. 4163.
17. M.A. Kastner, R.J. Birgeneau, G. Shirane, Y. Endoh *Rev. Mod. Physics* **70** (1998), p. 897.
18. Loram *et al. Proc. 10$^{th}$ Anniv. HTS Workshop*, (World Scientific, Singapore, 1996), p. 341.
19. D. Shoenberg *Superconductivity* (Cambridge University Press, Cambridge, 1954), p.164.
20. C. Panagopoulos *et al. Phys. Rev. Lett.* **79** (1997), p. 2320.
21. W. Barford and J.M.F. Gunn *Physica C* **156** (1988), p. 515.
22. R. Cywinski and B.D. Rainford *Hyperfine Interact.* **85** (1994), p. 215.
23. C. Panagopoulos *et al. Phys. Rev. B* **66** (2002), p. 064501.
24. P.M. Singer and T. Imai *Phys Rev Lett.* **88** (2002), p. 187601.
25. F. C. Chou, N.R. Belk, M.A. Kastner, R.J. Birgeneau and A. Aharony *Phys. Rev. Lett.* **75** (1995) p. 2204.
26. S. Wakimoto, S. Ueki, Y. Endoh and K. Yamada *Phys. Rev. B* **62** (2000), p. 3547.
27. M. Matsuda *et al. Phys. Rev. B* **65** (2002), p. 134515.
28. R. De Renzi *et al (these proceedings)*.
29. J.M. Tranquada *et al. Phys. Rev. Lett.* **78** (1997), p. 338.
30. V.J. Emery and S.A. Kivelson *J. Phys. Chem. Solids* **59** (1998), p. 1705.
31. J. Zaanen *J. Phys. Chem. Solids* **59** (1998), p. 1769.
32. C.M. Smith, A.H. Castro Neto and A.V. Balatsky *Phys. Rev. Lett.* **87** (2001), p. 177010.
33. H. Kimura *et al. Phys. Rev. B* **59** (1999), p. 6517.
34. M.-H. Julien *et al. Phys. Rev. Lett.* **84** (2000), p. 3422.
35. T. Xiang and J.M. Wheatley *Phys. Rev. Lett.* **77** (1996), 4632.
36. A. Ino *et al. J. Phys. Soc. Jpn.* **68** (1999), p. 1496.
37. C. Panagopoulos, J.R. Cooper, T. Xiang, Y.S. Wang and C.W. Chu *Phys. Rev. B* **61** (2000), p. R3808.
38. T. Xiang and C. Panagopoulos *Phys. Rev. B* **61** (2000), p. 6343.
39. W. Anukool, C. Panagopoulos and J.R. Cooper (*preprint*).
40. K.M. Lang *et al. Nature (London)* **415** (2002), p. 412.
41. Ch. Niedermayer *et al. Phys. Rev. Lett.* **71** (1993), p. 1764.
42. Y.J. Uemura *et al. Nature (London)* **364** (1993), p. 605.
43. J.L. Tallon, J.W. Loram and C. Panagopoulos, *MOS2002 J. Low Temp. Phys.* (in press).
44. S. Chakravarty, B.I. Halperin, and D.R. Nelson *Phys. Rev. Lett.* **60** (1988), p. 1057; S. Chakravarty, R.B. Laughlin, D.K. Morr and C. Nayak *Phys. Rev. B* **63** (2001), p. 10000.
45. C.M. Varma, P.B. Littlewood, S. Schmittrink, E. Abrahams and A.E. Ruckenstein *Phys. Rev. Lett.* **63** (1989), p. 1996; C.M. Varma *in Strongly Correlated Electronic Materials* Ed. K.S. Bedell, Addison Wesley (1994); C.M. Varma *Phys. Rev. B* **55** (1997), p. 14554; C.M. Varma *Phys. Rev. Lett.* **83** (1999), p. 3538.
46. S. Sachdev and J. Ye *Phys. Rev. Lett.* **69** (1992), p. 2411.
47. C. Castellani, C. DiCastro and M. Grilli *Phys. Rev. Lett.* **75** (1995), p. 4650.
48. D. Pines *Physica C* **341** (2000), p. 59.
49. A. Abanov, A.V. Chubukov and J. Schmalian *Europhys. Lett.* **55** (2001), p. 369.
50. A.V. Chubukov, D. Pines and J. Schmalian, Review Chapter to appear in *The Physics of Conventional and Unconventional Superconductors* edited by K.H. Bennemann and J.B. Ketterson (Springer-Verlag); Preprint cond-mat/0201140 at (http://xxx.lanl.gov) (2002).
51. Q. Si, S. Rabello, K. Ingersent and J. Smith Lleweilun *Nature (London)* **413** (2001), p. 804.

**Figure captions**

Fig. 1. Doping dependence of $T_c$ for the $La_{2-x}Sr_xCu_{1-y}Zn_yO_4$ ($y$=0.00, 0.01, and 0.02) series.

Fig. 2. Zero-field $\mu$SR spectra of $La_{2-x}Sr_xCuO_4$ for $x$=0.08 measured at different temperatures. The solid lines are the fits discussed in the text.

Fig. 3. Typical temperature dependence of the exponent $\beta$ of $La_{2-x}Sr_xCuO_4$ for $x$=$p$=0.08-0.17. The left-hand panel shows linear plots with an arrow showing $T_f$ for $p$=0.10, whereas as in the right-hand panel we show semi-log plots with the arrow indicating a $T_g$.

Fig. 4. Doping dependence of $T_g$ (closed circles), $T_f$ (open circles) and $T_c$ (crosses) of of $La_{2-x}Sr_xCuO_4$. The data for $p$<0.02 are from Ref. [13]. The inset is a semi-log plot of $T_g$ (mulitplied by 12) and $T_f$ as a function of doping.

Fig. 5. Doping dependence of $T_g$ (closed symbols) and $T_f$ (open symbols) of $La_{2-x}Sr_xCu_{1-y}Zn_yO_4$; $y$=0 (circles), $y$=0.01 (triangles) and $y$=0.02 (diamonds). The values of $T_c$ are shown as crosses for all values of $y$ (indicated in the figure).

Fig. 6. Doping dependence of $T_c$ for the $Bi_{2.1}Sr_{1.9}Ca_{1-x}Y_xCu_2O_{8+\delta}$ samples.

Fig. 7. The doping dependence of $T_g$ (closed symbols) and $T_f$ (open symbols) of $La_{2-x}Sr_xCu_{1-y}Zn_yO_4$; $y$=0 (circles), $y$=0.01 (upper triangles) and $y$=0.02 (diamonds) and $Bi_{2.1}Sr_{1.9}Ca_{1-x}Y_xCu_2O_{8+\delta}$ (lower triangles). Typical values for $T_c$ shown as crosses are for $La_{2-x}Sr_xCuO_4$. The broken line is the extrapolation of $T^*$ to T=0 [2].

Fig. 8. The upper panel shows the temperature dependence of the normalised superfluid density for grain-aligned $La_{2-x}Sr_xCuO_4$ at various dopings compared with the weak-coupling BCS theory (solid line) for a $d$-wave superconductor. The lower panel depicts the doping dependence of the absolute value of the superfluid density for the same material.

Fig. 9. Doping dependence of the inverse square of the absolute value of the $c$-axis penetration depth for two monolayer HTS, $La_{2-x}Sr_xCuO_4$ and $HgBa_2CuO_{4+\delta}$.

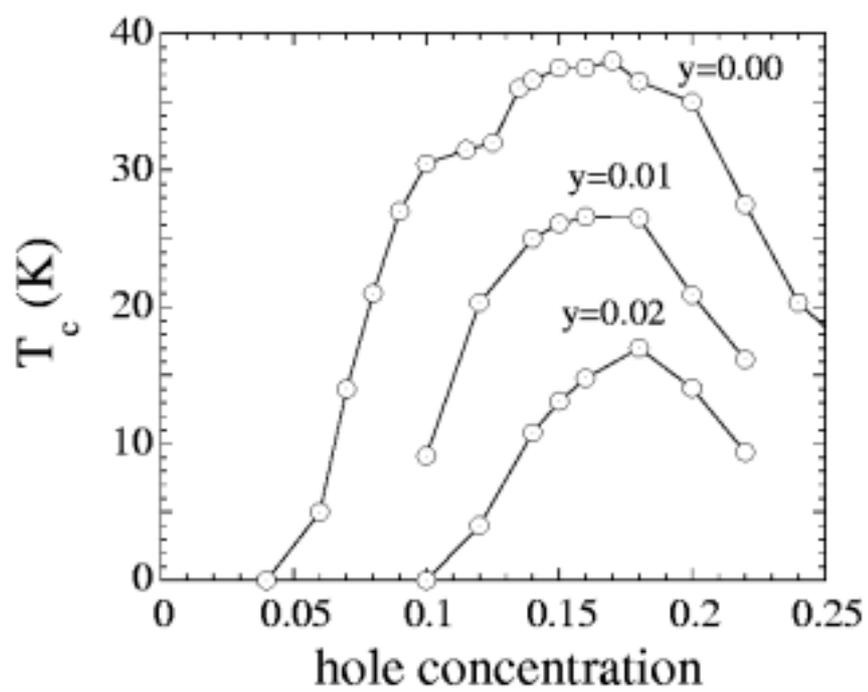

**FIG. 1**

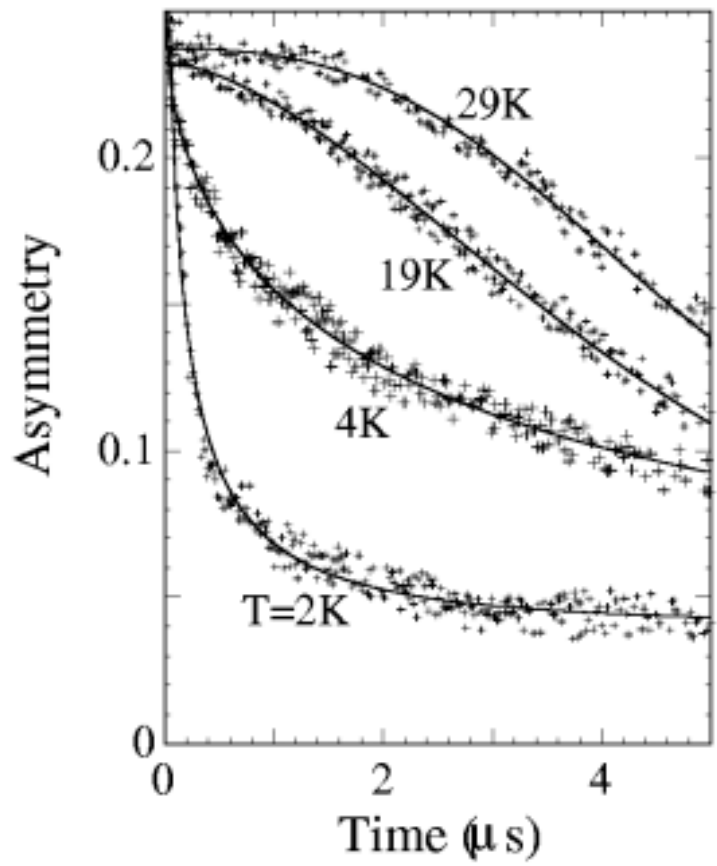

**FIG. 2**

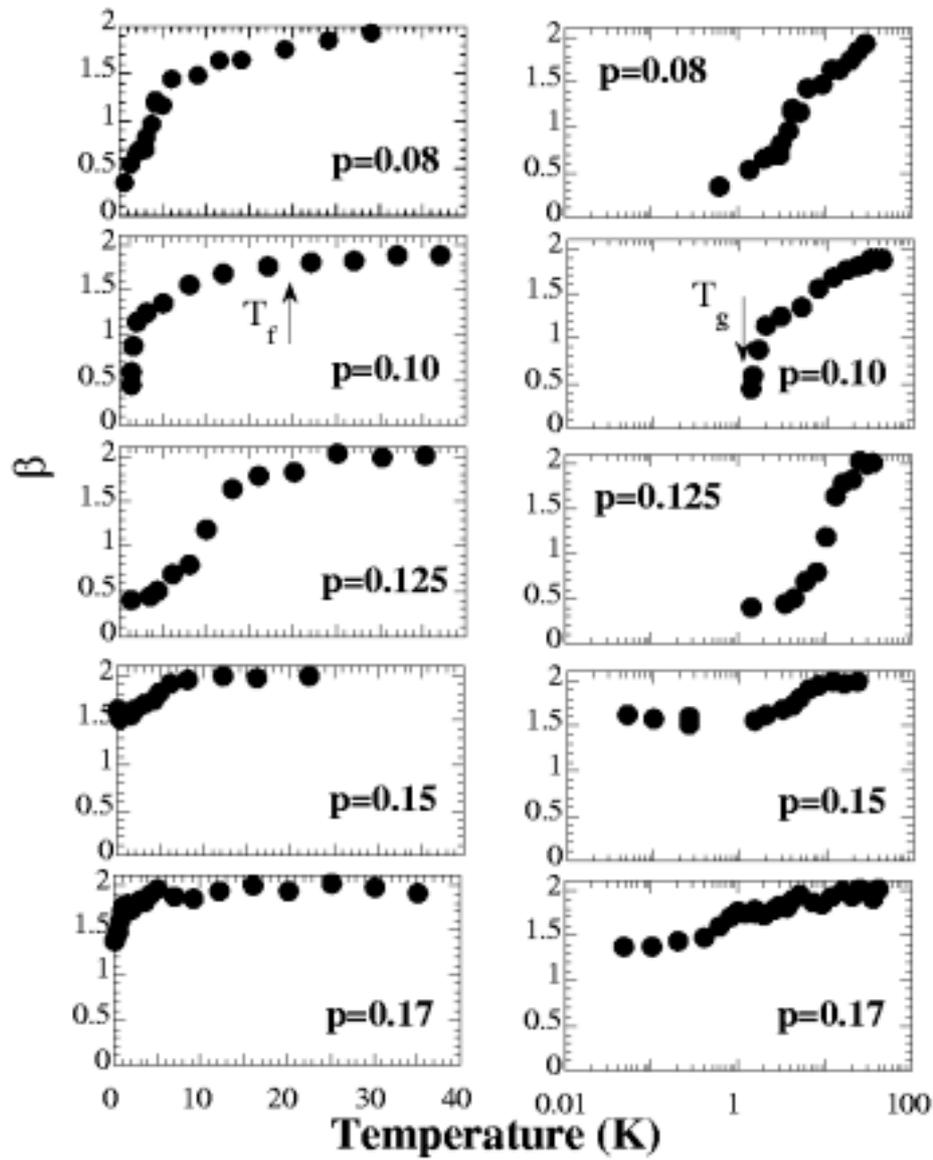

**FIG. 3**

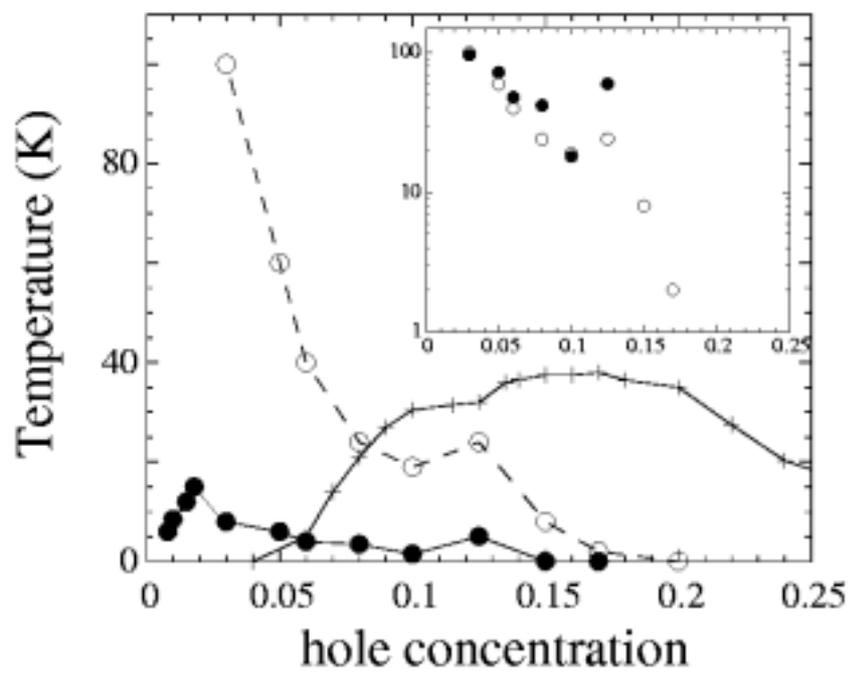

**FIG. 4**

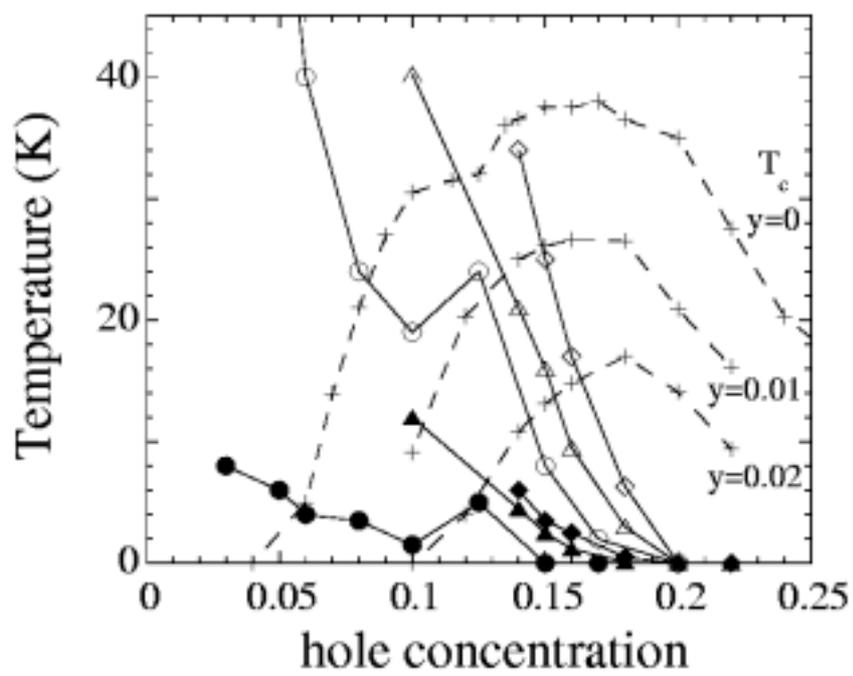

**FIG. 5**

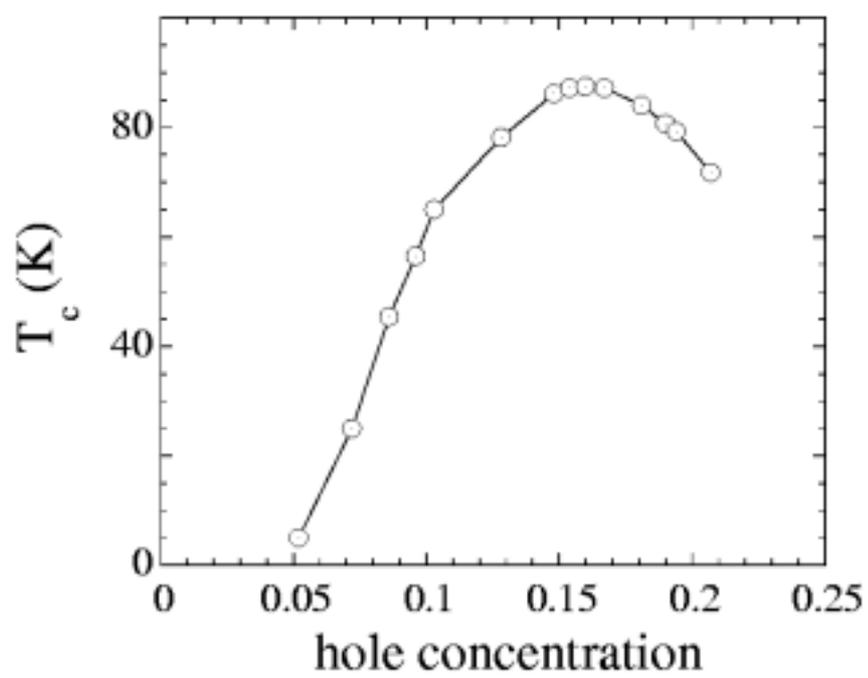

**FIG. 6**

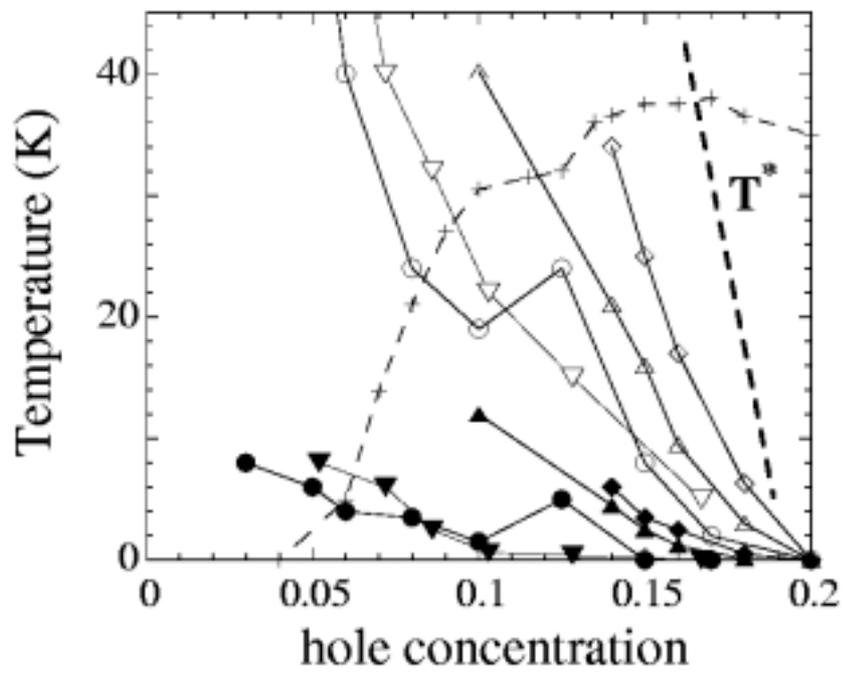

**FIG. 7**

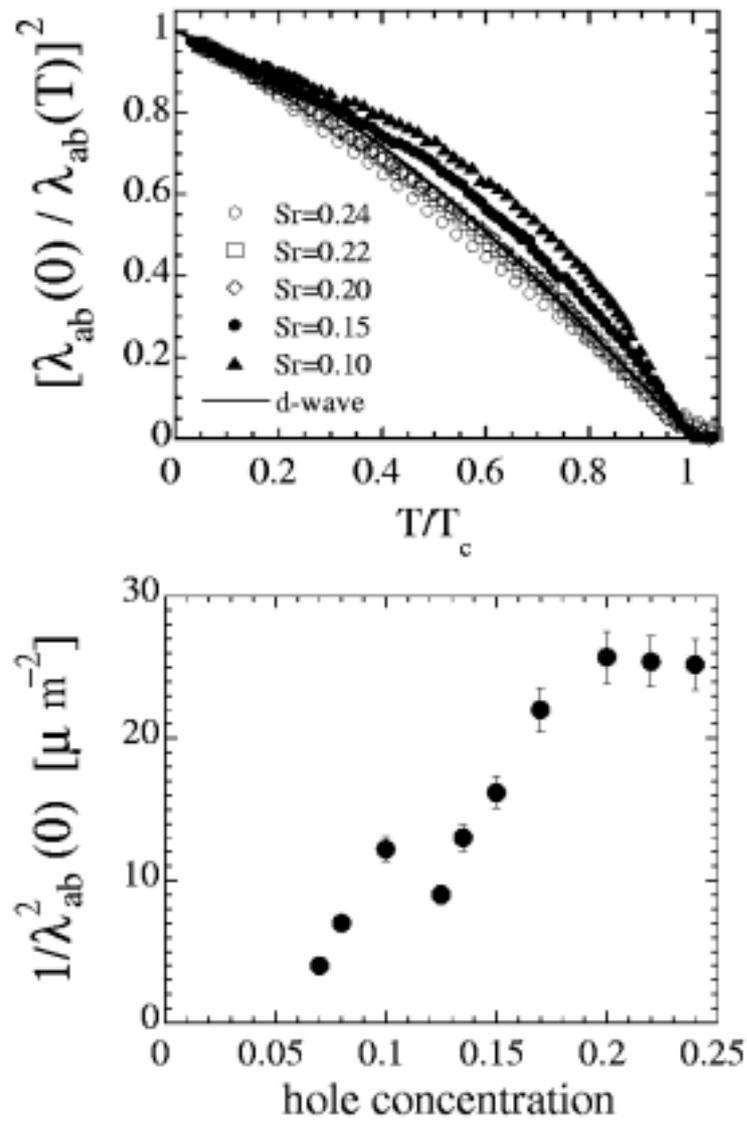

**FIG. 8**

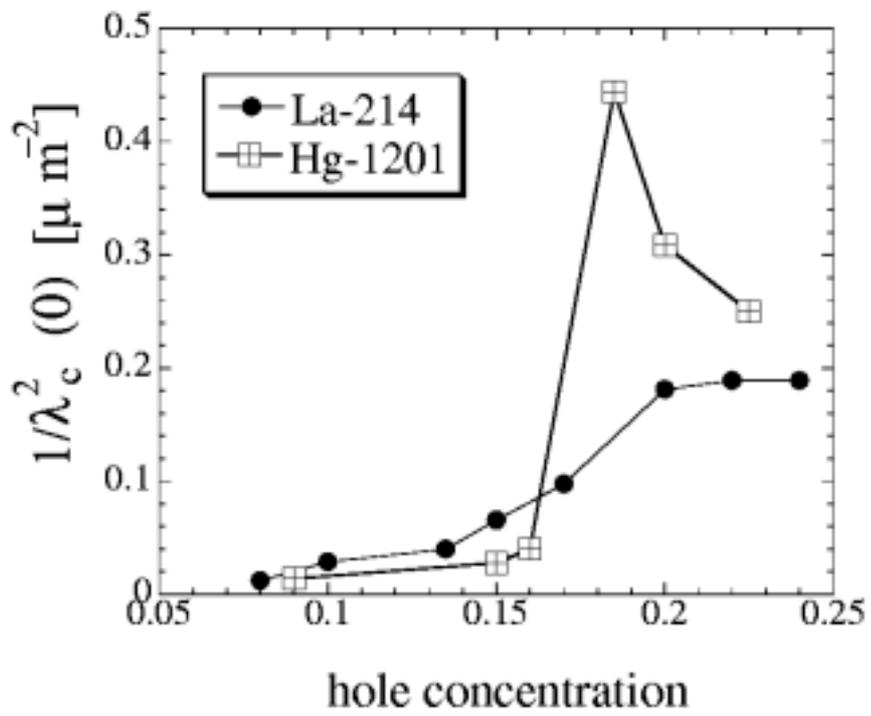

**FIG. 9**